\documentclass[fleqn]{wlscirep}
\usepackage{amsmath}

\newcommand{\erf}{\mathop{\mathrm{erf}}}
\DeclareMathOperator{\AIC}{AIC}

\title{The topology of large Open Connectome networks for the human
brain}

\author[1,2,*]{Michael T. Gastner}
\author[2]{G\'eza \'Odor}
\affil[1]{Yale-NUS College, 16 College Avenue West, \#01-220 Singapore 138527}
\affil[2]{MTA-EK-MFA, Research Center for Energy,
Hungarian Academy of Sciences, P. O. Box 49, H-1525 Budapest, Hungary}
\affil[*]{michael.gastner@yale-nus.edu.sg}

\begin{abstract}

The structural human connectome  (i.e.\ the network of fiber
connections in the brain) can be analyzed at ever finer spatial resolution
thanks to advances in neuroimaging.
Here we analyze several large data sets for the human brain network
made available by the Open Connectome Project.
We apply statistical model selection to characterize the degree
distributions of graphs containing up to $\simeq 10^6$ nodes and
$\simeq 10^8$ edges.
  A three-parameter generalized Weibull (also known as a stretched
  exponential) distribution is a good fit to most of the observed
  degree distributions.
  For almost all networks, simple power laws cannot fit the data, but
  in some cases there is statistical support for power
  laws with an exponential cutoff.
We also calculate the topological (graph) dimension $D$ and the
small-world coefficient $\sigma$ of these networks.
While $\sigma$ suggests a small-world topology, we found that $D < 4$
showing that long-distance connections provide only a small correction to the
topology of the embedding three-dimensional space.

\end{abstract}

\begin{document}

\maketitle

\section*{Introduction}

The neural network of the brain can be considered on structural (wired)
as well as on functional connection levels. Precise structural maps exist
only on very small scales. Functional networks, based on fMRI, are
available on larger datasets. However a clear-cut relationship between
them is largely unknown.
Drawing parallels between neural and socio-technological networks,
neuroscientists have hypothesized that in the brain we have small-world
networks, both on a structural~\cite{H2000} and functional
level~\cite{SpHo06}. Small-world networks are at the same time highly
clustered on a local scale, yet possess some long-distance connections that
link different clusters of nodes together.
This topology is efficient for signal processing~\cite{LF2000,
GMS12}, but doubts have remained if the small-world assumption is
generally true for the brain. Despite some evidence that functional networks
obtained from spatially coarse-grained parcellations of the brain are small
worlds~\cite{Salvador_etal05}, at a structural cellular level the brain
may be a large-world network after all~\cite{GMS12,CCH15}.

Like the small-world property, the hypothesis that functional brain networks have
scale-free degree distributions became popular around the turn of the
millennium~\cite{Egui-SF, Heu-SF}. The degree $k_i$ of node $i$ is defined as
the number of edges adjacent to $i$. Because the degree is a basic measure of a
node's centrality, the probability $\Pr(k)$ that a node has degree $k$ has played a
key role in network science for a long time~\cite{Solla76}. Especially physicists
have popularized power law fits to observed degree
distributions~\cite{BarabasiAlbert99, Caldarelli07}. When such a fit is
statistically justified, the network is called formally ``scale-free''. Power laws
play a crucial role in statistical physics, where they arise at transitions between
an ordered and unordered phase because of the absence of a characteristic length
scale. There are theoretical and empirical arguments that the brain operates near
such a critical point~\cite{BP03, Expert_etal11, ShewPlenz13, Haimovici_etal13}.
For this reason it is plausible to assume that, on a functional level, the
connectome's degree distribution is also scale-free. More sophisticated statistical
analyses of the functional connectome justify skepticism about the scale-free
hypothesis~\cite{Ferrarini_etal11, Le13, Ruiz14}. Until now there are only
few results for degree distributions of structural brain networks, 
and these do not show
clear evidence for power laws~\cite{Humphries_etal06}.

In this article we answer whether the structural connectome at an
intermediate spatial resolution can be viewed as a scale-free,
small-world network.
We analyze large data sets collected by the Open Connectome project
(OCP) \cite{OCP} that describe structural (rather than functional) brain connectivity.
The particular data sets chosen by us were processed by members of the
OCP from the raw diffusion tensor imaging data by Landman et al.~\cite{DTI}
Earlier studies of the structural network have analyzed much smaller data.
For example the network obtained by Sporns et al., using diffusion imaging
techniques \cite{29,30}, consists of a highly coarse-grained mapping
of anatomical connections in the human brain, comprising $N = 998$
brain areas and the fiber tract densities between them.
The entire brain is made up of $\sim 9\times 10^{10}$
neurons~\cite{Lent_etal11}, but current imaging techniques cannot
resolve such microscopic detail.
The networks investigated in this article have up to $\sim 10^6$ nodes,
which puts them on a scale that is halfway between the earlier
coarse-grained view and the complete neural network.

One important measure which could not have been estimated previously
because of too coarse-grained data is the topological (graph) dimension $D$.
It is defined by
\begin{equation} \label{topD}
N_r \sim r^D \ ,
\end{equation}
where $N_r$ is the number of node pairs that are at a topological
(also called ``chemical'')
distance $r$ from each other (i.e.\ a signal must traverse at
least $r$ edges to travel from one node to the other).
The topological dimension characterizes how quickly the whole network
can be accessed from any of its nodes:
the larger $D$, the more rapidly the number of $r$-th nearest
neighbors expands as $r$ increases.
Different small-world networks can possess different $D$, for example due to
their distinct clustering behavior~\cite{CCH15}.
Therefore, the level of small-worldness (quantified for example by the
coefficient defined by Humphries and Gurney~\cite{HumphriesGurney08}) and the
topological dimension contain different information.

Distinguishing between a finite and infinite topological dimension
is particularly important theoretically.
It has been conjectured that heterogeneities can cause strong
rare-region effects and generic, slow dynamics in a so-called
Griffiths phase~\cite{Griff}, provided $D$ is finite~\cite{Ma2010}.
Criticality~\cite{Obook} or even a discontinuous phase transition is smeared
over an extended parameter space.
As a consequence, a signalling network can exhibit behavior akin to
criticality, although it does not operate precisely on a unique
critical point that sharply divides an ordered from a disordered phase.
This phenomenon is pronounced for the Contact Process~\cite{harris74},
a common model for the spread of activity in a network.
Subsequent studies found numerical evidence for Griffiths effects
in more general spreading models in complex networks, although the scaling
region shrinks and disappears in the thermodynamic limit if
$D\to\infty$~\cite{BAGPcikk,wbacikk,basiscikk,ferrcikk}.
Recently Griffiths phases were also reported in synthetic brain
networks~\cite{MM,VMM,HMNcikk} with large-world topologies
and with modular organization, which enhances the capability
  to form
localized rare-regions.
Real connectomes have finite size, so they must possess finite $D$.
If in real connectomes $D$ remains small, these models hint at an
alternative explanation why the brain appears to be in a critical
state: instead of the
self-tuning mechanisms that have been frequently
postulated~\cite{Bak, HesseGross14}, the brain may be in a Griffiths phase, where
criticality exists without fine-tuning.
Models with self-tuning require two competing timescales: a slow "energy"
accumulation on the nodes and a fast redistribution by avalanches when the energy
reaches the firing threshold.
It is unclear if such a separation of timescales is realistic.
Even if the brain were in a self-organized critical state with clearly
separated timescales, Griffiths effects can
play an important role due to the heterogeneous behavior of the system,
frequently overlooked when modelling the brain.

\section*{Open Connectome brain network data}

The data sets analyzed in this article were generated by members of the OCP with the
MIGRAINE method described by Roncal et al.~\cite{MIG}
In this section we will briefly summarize their methods.
Afterwards we will describe our analysis which was based on the
graphs publicly available from the OCP web site~\cite{OCP}.
The raw input data used by the OCP consist of both diffusion and
structural magnetic resonance imaging scans with a resolution of
$\simeq 1\;\text{mm}^3$ (i.e.\ the size of a single voxel).
MIGRAINE combines various pieces of software into a ``pipeline'' to
transform this input to a graph with $10^5$--$10^6$ nodes.

As an intermediate step, the processing software first generates a
small graph of $70$ nodes~\cite{Gray}.
For this purpose the image is downsampled into $70$
regions taken from the Desikan gyral label
atlas~\cite{Desikan_etal06}.
During this step the software also identifies the fibers in the brain
with deterministic tractography using the Fiber Assignment by
Continuous Tracking (FACT) algorithm~\cite{Mori_etal99}.
As stopping thresholds a gradient direction of 70 degrees and a
stopping intensity of 0.2 were used.

These fibers are then reanalyzed in the next step of data processing.
The Magnetic Resonance One-Click Pipeline outlined by Mhembere et
al.~\cite{OCCres} generates a big graph where each voxel corresponds
to one node.
First a ``mask'' is defined, for example the 70 regions included in
the small graph.
Then all data outside the mask are discarded and
an edge is assigned to each remaining voxel pair that is connected by
at least one fiber staying within the boundaries of the mask.
This procedure will naturally produce hierarchical modular graphs with
(at least) two quite different scales.

At this point each scan has been turned into a network with $\simeq
10^7$ vertices and $\simeq 10^{10}$ edges.
However, due to the image processing algorithm
(especially because the mask is chosen conservatively),
many of these voxels will become disconnected and
must be considered as noise.
To clean up the data, all vertices outside the largest connected
component are removed.
According to Roncal et al.~\cite{MIG} the remaining graph ``keeps
essentially all white matter voxels, consisting of $\approx 10^5$
vertices and $\approx 10^8$ edges.''

One important point to note is that two voxels $A$ and $C$ are linked
by an edge even if there are other voxels $B_1,\ldots, B_n$ between
$A$ and $C$ on the same fiber.
For example, if one traverses voxels $A$, $B$, $C$ on a fiber, the
edges $(A,B)$, $(A,C)$ and $(B,C)$ are all part of the graph.
Furthermore, the edges are undirected so that $(B,A)$, $(C,A)$ and
$(C,B)$ are also part of the graph because the FACT algorithm cannot
provide information about the direction of an edge.
Note that confounding factors such as the measurement
technique, spatial sampling, measurement errors and the
network-construction method can affect the graph data
we downloaded \cite{Chaos10}. We cannot control them,
but tested the robustness of our conclusions by modifying one of
the networks by neglecting a fraction of edges that might have arisen
as a consequence of the transitivity rule. Additionally,
we tested the effect of changing the reference null model from
a nonspatial model (the Erd\H{o}s-R\'enyi graph) to a spatial one (the
random geometric graph,
see section ``Small-world coefficient'' below).

To save space the OCP data files store only one of the directions
(i.e.\ the upper triangle of the adjacency matrix) so that the
opposite direction must be inferred from the data and inserted into
the graph.

There were 3 different sets of big human brain graphs available from the
OCP website~\cite{OCP} with the abbreviations KKI (Kennedy Krieger Institute),
MRN (Mind Research Network) and NKI (Nathan Kline Institute).
The raw data are described by Landman et al.~\cite{DTI}, Jung et
al.~\cite{Jung_etal10} and Nooner et al.~\cite{Nooner_etal12}, respectively.
We analyzed the KKI graphs numbered 10 to 19 in more detail.
Some graph invariants (e.g.\ degrees, clustering coefficients) were
calculated and analyzed by Mhembere et al.~\cite{OCCres}, but for the
present study we have recalculated all invariants directly from the
graph data available from the OCP website.

\section*{Degree distribution}

\subsection*{Model selection}

\begin{figure}[h]
\includegraphics[height=5.5cm]{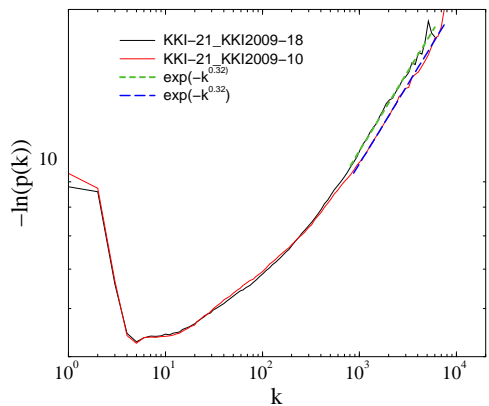}
\caption{\label{eloLpfit} Empirical degree distributions of the
KKI-10 and KKI-18 graphs
(solid lines).
Dashed lines show a rough-and-ready approach:
ordinary least-squares fits for $k>1000$ for the two graphs
(short-dashed line for fit to KKI-18, long-dashed line for KKI-10)
suggest stretched exponential tails.
While ordinary least-squares fits are not a sound basis for model
selection, we demonstrate in this article that there is indeed
statistical evidence in favor of a generalized three-parameter Weibull distribution
with a stretched exponential tail.}
\end{figure}

We want to assess how well different probability distributions fit the
degrees of the OCP graphs.
A first rough-and-ready visual attempt supports the hypothesis that
the tails might be stretched exponentials, for example for the
networks KKI-10 and KKI-18 in Fig.~\ref{eloLpfit}.
However, such visual techniques have no inferential power.
Even if the fitted parameters come from ordinary least-squares
regression toolboxes, the fitted parameters in general do not
converge to the true values even if the number of data points is large.
Moreover, least-squares regression lacks a statistically principled
criterion for comparing different models with each other.

A statistically sound framework for model selection is information theory.
Here we adopt the information-theoretic methodology proposed by
Handcock and Jones~\cite{HandcockJones04} who fitted different functions to
degree distributions in sexual contact networks.
The key idea is that a good model should perform well in terms of two
opposing objectives.
On one hand the model should have enough flexibility so that it is
able to fit the observed distribution.
On the other hand it should have only a minimal number of
parameters.
In general, the more parameters we have at our disposal, the better we
can fit the observation.

The goodness of fit can be quantified by the likelihood function
which, under the assumption of independent observations, has the form
\begin{align}
  \mathcal{L}(\mathbf{v}) =
  \prod_{i=1}^N\Pr(k_i,\mathbf{v}).
\end{align}
Here $\mathbf{v}$ is the set of parameters in the model, $N$ the
number of observations, and $k_1,\ldots,k_N$ are the observed degrees.
Alternatively we can write the likelihood as
\begin{align}
  \mathcal{L}(\mathbf{v}) = \prod_{i=k_\text{min}}^{k_\text{max}} \left[\Pr(k_i,\mathbf{v})\right]^{n_i},
\end{align}
where $k_\text{min}$ and $k_\text{max}$ are the minimum and maximum
observed degrees and $n_i$ is the number of times we observe the
degree $k_i$.
In the graphs KKI-10 through KKI-19, $k_\text{min}$ is always equal to $1$;
$k_\text{max}$ ranges from $5154$ to $11\,241$.
In the extreme case of allowing as many parameters as there are observed degrees
we can achieve a maximal likelihood of $e^{-nH}$,
where $H = -\sum_{i=k_\text{min}}^{k_\text{max}}
\frac{n_i}N\ln\left(\frac{n_i}N\right)$ is the Shannon entropy of
the data.
However, such a highly parameterized model is no longer informative,
because it fits only the particular connectome used as input and sheds
little light on general features that different connectomes might have
in common.

Statisticians have proposed several ``information criteria'' to
address this problem of over-fitting (e.g.\ Bayesian, deviance or
Hannan-Quinn information criteria).
These are objective functions that rate the quality of a model based
on a combination of the likelihood $\mathcal{L}$ and the number of
parameters $K$.
In this study we apply the Akaike information criterion with a
correction term for finite sample size~\cite{BurnhamAnderson98},
\begin{align}
  \AIC_c = -2\ln(\mathcal{L}(\hat{\mathbf{v}})) + 2K +
  \frac{2K(K+1)} {N-K-1},
  \label{AICc}
\end{align}
where $\hat{\mathbf{v}} = (\hat{v}_1,\ldots,\hat{v}_K)$ is the set of
parameters that maximizes $\mathcal{L}$; as before, $N$ is the number
of observations.
The last term in Eq.~\ref{AICc} is a second-order bias correction
which, although not in Akaike's original formula~\cite{Akaike74},
gives more accurate results if $N\simeq K$~\cite{HurvichTsai89}.

While the absolute size of $\AIC_c$ is not interpretable, differences
in $\AIC_c$ are informative and allow us to quantitatively compare
different models~\cite{BurnhamAnderson98}.
If we denote the $\AIC_c$ value of model $j$ by $\AIC_c^{(j)}$ and the
minimum over all models by $\AIC_c^{(\text{min})}$, then the difference
\begin{equation}
  \Delta_j = \AIC_c^{(j)} - \AIC_c^{(\text{min})}
  \label{Delta_j}
\end{equation}
estimates the relative expected Kullback-Leibler distance between
model $j$ and the estimated best model~\cite{BurnhamAnderson01}.
As a rule of thumb, models with $\Delta_j\lesssim2$ have substantial
empirical support; models with $\Delta_j>10$ on the other hand are
unlikely candidates to explain the data~\cite{BurnhamAnderson98}.

Burnham and Anderson~\cite{BurnhamAnderson98} list many theoretical reasons in
favor of model selection based on $\AIC_c$.
The Bayesian information criterion (BIC), although almost equally
popular, has been reported to yield poor results when used to fit
power-law tails in probability distributions~\cite{Clauset_etal09}
because it tends to underestimate the number of parameters.
The Akaike information criterion penalizes less severely for
additional parameters: in the limit $N\gg K$ the penalty is
asymptotically equal to the term $2K$ in Eq.~\ref{AICc}, whereas the
equivalent term in the BIC grows as $K\ln(N)$.
We carried out Monte Carlo simulations on synthetically
generated probability distributions of the type described in the
next section (see supplementary material).
We found that model selection by $\AIC_c$ came close to
the true number of parameters.
Although the BIC showed acceptable performance, we confirmed that it indeed
favors a too small number of parameters.
We therefore advocate the use of $\AIC_c$ rather than BIC for fitting
degree distributions.

\subsection*{Candidate models}
The first step of $\AIC_c$-based model selection is the definition of
several candidate models that might generate the observed
distribution.
We denote as before by $\Pr(k)$ the probability that a node has degree $k$.
The distinctive feature of different candidate models is the
asymptotic decay of $\Pr(k)$ for $k\gg 1$.
Only in this limit we can hope to find scale-free behavior if it
indeed exists.
Of course, all real networks are finite so that, strictly speaking,
we cannot take the limit $k\to\infty$.
If we restrict ourselves to only a few high-degree nodes, we have too few data points
for a meaningful fit.
On the other hand, if we include too many low-degree nodes, then we
may misjudge the correct asymptotic behavior of $\Pr(k)$.

We therefore assume for all candidate models that there is an optimal
cutoff point $k_c$ that separates nodes  with degree $\leq k_c$ from
the region where a hypothesized asymptotic function $F(k)$ can fit the
data~\cite{HandcockJones04},
\begin{equation}
  \Pr(\text{degree} > k) =
  \begin{cases}
    1-\sum_{i=1}^k A_i & \text{if $k \leq k_c$}\\
    \frac{1-\sum_{i=1}^{k_c}A_i}{F(k_c)}F(k) & \text{if $k > k_c$}.
  \end{cases}
\end{equation}
Each parameter $A_k$ is chosen so that $\Pr(k)=A_k$ for $k=1,\ldots,k_c$.
Different families of candidate models can be defined by different functions $F(k)$.
We list all the functions $F$ investigated in this study in
Table~\ref{candidates}.
The exponential function (EXP) is the candidate that decays
most rapidly in the right tail.
The power law (POW) has two parameters: an exponent $\beta>0$ and a
constant $\alpha>0$ that shifts the function to the left or
right. The tail $k\gg\alpha$ has the conventional power law form
$F(k)\propto k^{-\beta}$.
The discrete log-normal (LGN) and Weibull (WBL) distributions are
represented by the usual distribution functions of their continuous
namesakes.
We also include two three-parameter models: a truncated power law (TPW)
and the generalized Weibull distribution (GWB).
In comparison to POW, TPW includes an additional exponential factor
which is often used to mimic finite-size cutoffs in the right tail.
GWB is a standard three-parameter generalization of WBL with an
additional parameter $\gamma$, called location parameter, that shifts
the distribution to the right or left~\cite{TeimouriGupta13}.

\begin{table}
  \renewcommand{\arraystretch}{1.3}
  \centering
  \begin{tabular}{|c||c|}
    \hline
    model & $F(k)$ \\
    \hline
    \hline
    exponential (EXP) & $e^{-\alpha k}$ \\
    \hline
    power law (POW) & $\alpha^\beta(k+\alpha)^{-\beta}$ \\
    \hline
    log-normal (LGN) & $\frac12-\frac12\erf\left(\frac{\ln
                k-\alpha}{\sqrt2\beta}\right)$ \\
    \hline
    Weibull (WBL) & $\exp\left(-\alpha k^\beta\right)$ \\
    \hline
    truncated power law (TPW) & $\alpha^\beta(k+\alpha)^{-\beta}e^{-\gamma
                          k}$ \\
    \hline
    generalized Weibull (GWB) &
                          $\exp\left[\alpha\left(\gamma^\beta-(k+\gamma)^\beta\right)\right]$\\
    \hline
  \end{tabular}
  \caption{Investigated candidate models for the degree distribution.}
  \label{candidates}
\end{table}

We can distinguish the members of each candidate model family by the choice of
$k_c$, the values of $A_1,\ldots, A_{k_c}$ and $\alpha$, $\beta$, $\gamma$.
Model selection by $\AIC_c$ gives a natural criterion for the
optimal parameters matching an observed distribution.
It is a simple exercise to prove that $\mathcal{L}$ is maximized if
$A_k$ equals the observed relative frequency of nodes with degree $k$,
\begin{equation}
  A_k = \frac{n_k}N.
\end{equation}
For a fixed value of $k_c$, standard numerical algorithms can optimize
the remaining parameters $\alpha$, $\beta$ and $\gamma$ in
Table~\ref{candidates} to maximize $\mathcal{L}$.
After calculating these maximum-likelihood estimators for every $k_c$
between $0$ and $k_\text{max}$, we search for the value of $k_c$ that
minimizes $\AIC_c$ of Eq.~\ref{AICc}. The number $K$ of parameters
that we have to insert into this equation is
\begin{itemize}
  \item $K = k_c+1$ for EXP,
  \item $K = k_c+2$ for POW, LGN and WBL,
  \item $K = k_c+3$ for TPW and GWB.
\end{itemize}

\subsection*{Results of model selection}
\begin{figure}
  \includegraphics[height=5.5cm]{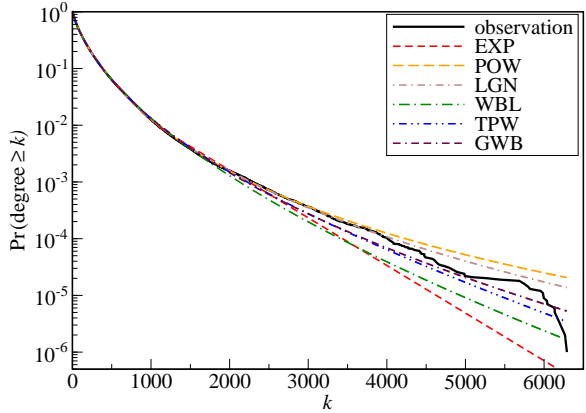}
  \caption{\label{ccdf_fit} The maximum-likelihood distributions from
    each model family for matching the degree distribution of network KKI-18. In this example the
    generalized Weibull distribution is the best compromise in the
    right tail (see Table~\ref{AIC_table}).}
\end{figure}

For each candidate model and each $k_c=0, \ldots,
k_\text{max}$  we compute the $\AIC_c$.
We then determine the smallest $\AIC_c^\text{(min)}$ from the entire
set and calculate for each model $j$ the difference $\Delta_j$ defined
in Eq.~\ref{Delta_j}.
In Fig.~\ref{ccdf_fit} we show for the example of the network KKI-18 the best-fitting
distribution within each candidate model family.
We have chosen a logarithmic scale for the ordinate to highlight the
differences in the right tail.
While the exponential and simple Weibull distributions decrease too rapidly, the
power law and log-normal distributions decay too slowly.
The truncated power law and generalized Weibull distributions are
better in mimicking the overall shape of the distribution.

We find that, broadly speaking, this observation is typical of the ten
investigated connectomes.
As we see from Table~\ref{AIC_table}, in 9 out of the 10 investigated networks we
can achieve $\Delta_j<10$ with the GWB distribution.
In 6 out of 10 cases, there is a similar level of evidence for the TPW
model which has been previously hypothesized for functional brain networks~\cite{4,63,64}.
For all other candidate distributions there is either none or very
sporadic statistical support.
When interpreting Table~\ref{AIC_table}, one should bear in mind
that even the best-fitting model is only ``best''  compared with the other
tested candidates.
True brain networks are of course far more complicated than any of
our candidate models.
Even if there were a ``true'' model, searching for ``the'' degree
distribution of the human connectome is not a sensible endeavor
because it would certainly be a highly complex model that is unlikely
ever to be discovered and included in the set of candidates.
As we discuss in the supplementary information,
we can nevertheless sensibly ask
which candidate model comes closest to the truth in the sense
that it minimizes the Kullback-Leibler divergence.
With this interpretation, we conclude that GWB is generally the best
of our candidates.
The fact that not all of the ten data sets are best fitted by the same
model does not call this conclusion into question.
Just as in traditional $p$-value based hypothesis testing, we also
expect in $\AIC_c$-based model selection that the best general model is
sometimes rejected by random chance for a concrete data sample.

\begin{table}
  \renewcommand{\arraystretch}{1.3}
  \centering
  \begin{tabular}{|c||c|c|c|c|c|c||c|c|c|c|c|c|}
    \hline
    connectome & \multicolumn{6}{c||}{$\Delta_j$}\\
    KKI-$\ldots$ & EXP & POW & LGN & WBL & TPW & GWB\\
    \hline
    \hline
    $10$ & $1317.82$ & $202.90$ & $204.81$ & $73.70$ & $\mathbf{0.00}$
                                         & $\mathbf{3.52}$\\
    \hline
    $11$ & $1245.42$ & $17.39$ & $\mathbf{0.00}$ & $38.75$ &
                                                             $\mathbf{1.21}$ & $\mathbf{0.43}$\\
    \hline
    $12$ & $992.62$ & $43.28$ & $21.83$ & $174.19$ & $\mathbf{0.00}$ &
                                                                       $\mathbf{0.73}$\\
    \hline
    $13$ & $767.95$ & $155.43$ & $72.68$ & $221.50$ & $\mathbf{3.45}$
                                         & $\mathbf{0.00}$\\
    \hline
    $14$ & $792.10$ & $117.26$ & $124.21$ & $85.74$ & $\mathbf{0.00}$
                                         & $\mathbf{9.57}$\\
    \hline
    $15$ & $1094.40$ & $120.99$ & $139.87$ & $26.19$ & $86.06$ &
                                                                 $\mathbf{0.00}$\\
    \hline
    $16$ & $954.28$ & $18.81$ & $29.99$ & $66.59$ & $\mathbf{0.00}$ & $17.20$\\
    \hline
    $17$ & $736.42$ & $195.15$ & $216.05$ & $18.16$ & $49.47$ & $\mathbf{0.00}$\\
    \hline
    $18$ & $1168.50$ & $109.51$ & $109.90$ & $185.59$ & $51.81$ &
    $\mathbf{0.00}$\\
    \hline
    $19$ & $1006.80$ & $\mathbf{3.91}$ & $47.07$ & $190.19$ & $\mathbf{0.56}$ & $\mathbf{0.00}$\\
    \hline
  \end{tabular}
  \caption{ \label{AIC_table} Smallest relative Akaike information criterion
    $\Delta_j=\AIC^{(j)}_c-\AIC_c^\text{(min)}$ for each
    candidate degree distribution. We highlight in bold font all
    $\Delta_j<10$. For all other models $j$ there is essentially no
    empirical support~\cite{BurnhamAnderson98}.
  }
\end{table}

A closer look at the fitted GWB values (Table~\ref{tabgwb})
shows that, with the exception of KKI-16 where GWB is rejected by the $\AIC_c$, the
$\beta$ exponents lie within one order of magnitude suggesting a common
trend if not even universality.
This order of magnitude also agrees with the least-squares fit in Fig.~\ref{eloLpfit}.

\begin{table}
\begin{center}
\begin{tabular}{|r|r|r|r|r|}
\hline
id & $k_{c}$ & $\alpha$ & $\beta$ & $\gamma$ \\
\hline
10 & 115 & 0.239 & 0.4486 & 110.22 \\
11 & 309 & 1.619 & 0.2700 & 208.37 \\
12 & 119 & 2.597 & 0.2382 & 270.62 \\
13 & 35  & 0.669 & 0.3519 & 100.98 \\
14 & 114 & 0.411 & 0.4232 & 199.25 \\
15 & 83  & 0.064 & 0.6021 & -18.51 \\
16 & 101 & 45.670 & 0.0507 & 379.24 \\
17 & 21  & 0.057 & 0.6437 & -4.60 \\
18 & 94  & 0.358 & 0.4260 & 141.72 \\
19 & 112 & 8.329 & 0.1645 & 502.64 \\
\hline
\end{tabular}
\caption{\label{tabgwb} Summary of fitted parameters of the GWB model
  for the investigated KKI graphs.}
\end{center}
\end{table}

\section*{Dimension measurements}

\begin{figure}[h]
\includegraphics[height=5.5cm]{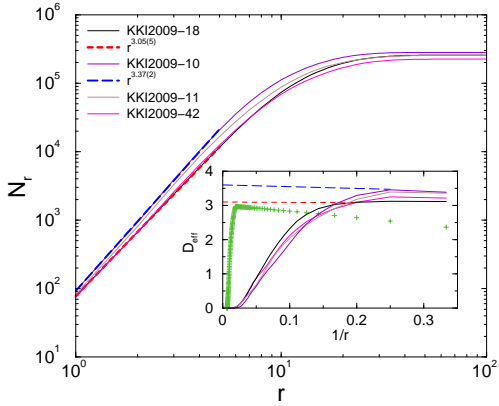}
\caption{\label{dim-KKI}Number of nodes within graph distance $r$
in the big KKI graphs. Dashed lines show power-law fits. Inset: local
slopes defined in Eq.~\ref{Deff}.
Crosses correspond to measurements on a regular $100^3$ lattice.}
\end{figure}

To measure the dimension of the network~\cite{GastnerNewman06} we first computed the
distances from a seed node to all other nodes by running the
breadth-first search algorithm.
Iterating over every possible seed, we counted the number of nodes
$N_r$ with graph distance $r$ or less from the seeds and calculated
the averages over the trials.
As Fig.~\ref{dim-KKI} shows, an initial power law crosses over
to saturation due to the finite network sizes.
We determined the dimension of the network, as defined by the scaling
law (\ref{topD}),
by attempting a power-law fit to the data $N(r)$ for the initial
ascent.
This method resulted in dimensions between $D=3$ and $D= 4$.

To see the corrections to scaling we determined the effective exponents
of $D$ as the discretized, logarithmic derivative of (\ref{topD})
\begin{equation}  \label{Deff}
D_\mathrm{eff}(r+1/2) = \frac {\ln N_r - \ln N_{r+1}} {\ln(r) - \ln(r+1)} \ .
\end{equation}
As the inset of Fig.~\ref{dim-KKI} shows, $D_\mathrm{eff}(r)$ tends to
values below $4$ even in the infinite size limit, but the narrow
scaling region breaks down for $r > 5$. Furthermore, the extrapolated
values for $D$ of the connectomes exhibit an increasing tendency with $N$ as we will
now explain in detail.

\begin{figure}[h]
\includegraphics[height=5.5cm]{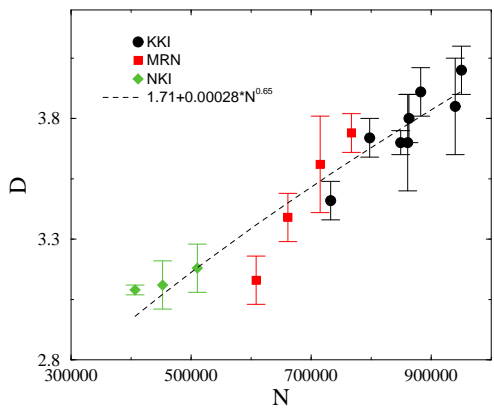}
\caption{\label{dim-par} Topological dimension as a function of network
size using different data sets. The line shows a power-law fit for the combined
KKI, MRN, NKI results.}
\end{figure}

For better understanding we have also performed the analysis for other graphs
besides KKI, possessing graph sizes in different ranges of $N$.
The finite size scaling results are summarized in Fig.~\ref{dim-par},
where we also extrapolate to $N\to\infty$. As one can see, the
dimension values follow the same trend for KKI-, MRN- and NKI-graphs
without any clear sign of saturation.
A power-law fit to the data with the form $A+B N^C$ is also shown,
suggesting that $D$ diverges for infinite $N$. It is tempting to
extrapolate with this function to larger sizes or even to the infinite
size limit.
However, since we can rule out that the degree distributions are
scale-free, we cannot assume that such extrapolated graphs
faithfully represent connectomes of finer resolution.
For example,
using this power-law extrapolation we would
overestimate
their maximum degree $k_\text{max}$.

Thus the present data does not permit claiming any particular numeric
value for the dimension $D$ of the true (i.e.\ microscopically
resolved) brain connectome.

 We cannot modify the algorithms that generate the OCP graphs,
but tested the robustness by randomly removing $20\%$ of the
directed graph
connections in case of the KKI-18 network. This makes the network
partially directed, more similar to a real connectome. As a
result the graph dimension did not change much: $D=3.2(2)$ instead of
$3.05(1)$.
Since the majority of edges are short, a random removal results in
a relative enhancement of long connections, but $D$ increases only slightly.

\section*{Small-world coefficient}

Small-worldness can be characterized by different definitions.
One of them is the so-called small-world
coefficient $\sigma$, which is defined as the normalized
clustering coefficient ($C/C_r$) divided by the normalized
average shortest path length ($L/L_r$),
\begin{equation}
\sigma = \frac{C/C_r}{L/L_r} \ ,
\label{swcoef}
\end{equation}
where the normalization divides the observed quantity ($C$ or $L$) by
the expectation value ($C_r$ or $L_r$) for an Erd\H{o}s-R\'enyi (ER) random
graph with the same number of nodes and edges~\cite{HumphriesGurney08}.

There are two different definitions of a clustering coefficient in the
literature.
The Watts-Strogatz clustering coefficient \cite{WS98} of a
network of $N$ nodes is
\begin{equation}\label{Cws}
C^{W} = \frac1N \sum_i 2n_i / k_i(k_i-1) \ ,
\end{equation}
where $n_i$ denotes the number of direct edges interconnecting the
$k_i$ nearest neighbors of node $i$.
An alternative is the ``global'' clustering coefficient \cite{globalC},
also called ``fraction of transitive triplets'' in the social networks
literature~\cite{Jackson08},
\begin{equation}\label{Cg}
C^{\Delta} = \frac{\rm number \ of \ closed \ triplets}
{\rm number \ of \ connected \ triplets} \ .
\end{equation}
Both definitions are in common use, but values for $C^W$ and
$C^{\Delta}$ can differ substantially because Eq.~\ref{Cws} gives
greater weight to low-degree nodes.

The average shortest path length is~\cite{Heuvel_etal09}
\begin{equation}
L = \frac{1}{N (N-1)} \sum_{j\ne i} d(i,j) \ ,
\end{equation}
where $d(i,j)$ is the graph distance between vertices $i$ and $j$.
$L$ is only properly defined if the network is connected because otherwise
the graph distance between some voxels is infinite.

\begin{figure}[h]
\includegraphics[height=5.5cm]{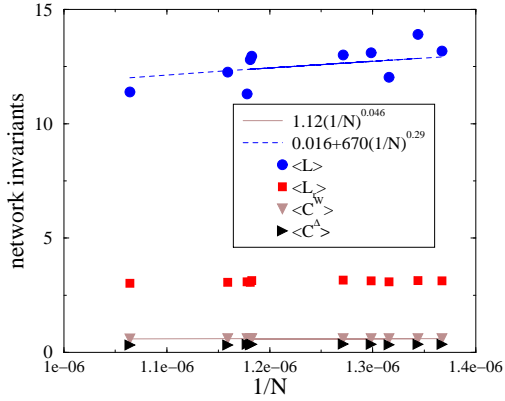}
\caption{\label{tab3} Network invariants as a function of $1/N$
for the investigated KKI graphs from Table~\ref{tab1}.
Lines show power-law fits.}
\end{figure}

We have calculated $C$ and $L$ for the largest connected components of several
KKI networks directly from the edge lists on the OCP website~\cite{OCP}.
The $C^{\Delta}$ values are about half of those for $C^{W}$ (see Table~\ref{tab1}).
A finite size scaling analysis shows that the values of $L$,
$C^{\Delta}$ and $C^{W}$ decay by increasing the size $N$ (see Fig.~\ref{tab3}).
For the average shortest path-length this decay is rather fast;
a least-squares regression with the form $a+b(1/N)^c$ results in
$c \simeq 0.29$ and $a\simeq 0.016$. The constant is near zero within
the precision of data, so in the infinite size limit we see small-world behavior.
The clustering coefficients are almost constant; the power-law fit provides
a very small slope: $c \simeq 0.046$ in agreement with the behavior of modular graphs
\cite{RB03}.
The decreasing trends for $L$, $C^W$ and $C^\Delta$ as functions of
$N$ are statistically significant at the 5\% significance level; the $p$-values for
$t$-tests of zero slope for the log-transformed data are $0.03$,
$0.002$ and $0.04$ respectively.
Again, finite size scaling has to be interpreted with caution given that the
topology is not scale-free.

As before,
we tested the robustness,
this time
by deleting $10\%$ randomly chosen undirected edges.
We obtained very
little changes:
$L=11.37$ (previously: $11.30$), $C^W = 0.538$ (previously: $0.598$)
$C^\Delta = 0.322$ (previously: $0.358$).

Due to the transitivity of OCP graphs, one may question the validity
of using ER graphs as a null model.
Especially the high clustering can be at least partly attributed to
the spatial embedding so that, for example, three-dimensional random
geometric graphs~\cite{Penrose03} are useful as comparison.
Random geometric graphs have indeed a much higher clustering
coefficient~\cite{RGG}, $C^W=15/32$, a similar value to $C^{W}$ in the
OC graphs.
To circumvent such problems Telesford et al. \cite{Tele} suggested another
small-world criterion, using a "latticized" version of the graph for reference.
However, as they pointed out, the latticization algorithm is
computationally too demanding for large graphs.
Their algorithm is in practice feasible only for up to $N\simeq 10^4$
nodes, thus the necessary memory and run-times are prohibitive in our
case.

We therefore kept the ER graphs as our null model and calculated the
corresponding small-world coefficient $\sigma$ in Eq.~\ref{swcoef}.
We determined the clustering coefficient of the
corresponding random networks
$C_r = \langle k\rangle / N$, where $\langle k\rangle$ is the
mean degree.
We have computed the average path length of the corresponding Erd\H{o}s-R\'enyi
networks with the formula~\cite{Fron}
\begin{equation}
L_r = \frac{\ln(N_l) - 0.5772}{\ln\langle k\rangle} + 1/2 \ ,
\end{equation}
where $N_l$ is the size of the largest component.

Applying these formulas to the KKI graphs, the last two columns of
Table~\ref{tab1} show that $\sigma^W,\sigma^\Delta \gg 1$ for all
cases, suggesting small-world behavior according to this definition.
These values do not show any tendency with respect to $N$:
the $t$-tests for zero slope have $p$-values $0.45$ and $0.72$ for
$\sigma^W$ and $\sigma^\Delta$, respectively.

\begin{table*}
\begin{center}
\begin{tabular}{|r|r|r|r|r|r|r|r|r|r|r|}
\hline
KKI & $N$ & $N_\text{edges}$ & $\langle k\rangle$ & $L$ & $L_r$ & $C^{W}$ & $C^{\Delta}$ & $C_r$ & $\sigma^W$ & $\sigma^\Delta$\\
\hline
10 & $9.40\times10^5$ & $8.68\times10^7$ & 184.71 & 11.38 & 3.02 & $5.94\times10^{-1}$ & $3.20\times10^{-1}$ & $1.97\times10^{-4}$ & 803.26 & 433.09 \\
11 & $8.63\times10^5$ & $7.07\times10^7$ & 163.84 & 12.25 & 3.07 & $5.99\times10^{-1}$ & $3.24\times10^{-1}$ & $1.90\times10^{-4}$ & 789.23 & 427.69\\
12 & $7.44\times10^5$ & $4.98\times10^7$ & 133.79 & 13.91 & 3.14 & $6.02\times10^{-1}$ & $3.58\times10^{-1}$ & $1.80\times10^{-4}$ & 757.12 & 450.43\\
13 & $8.46\times10^5$ & $5.93\times10^7$ & 140.17 & 12.96 & 3.14 & $6.02\times10^{-1}$ & $3.56\times10^{-1}$ & $1.66\times10^{-4}$ & 881.74 & 521.58\\
14 & $7.70\times10^5$ & $5.36\times10^7$ & 139.10 & 13.10 & 3.13 & $6.01\times10^{-1}$ & $3.62\times10^{-1}$ & $1.81\times10^{-4}$ & 794.64 & 478.99\\
15 & $8.47\times10^5$ & $6.94\times10^7$ & 163.84 & 12.80 & 3.06 & $5.99\times10^{-1}$ & $3.32\times10^{-1}$ & $1.94\times10^{-4}$ & 740.79 & 411.13\\
16 & $7.60\times10^5$ & $5.70\times10^7$ & 150.11 & 12.03 & 3.09 & $6.02\times10^{-1}$ & $3.38\times10^{-1}$ & $1.98\times10^{-4}$ & 782.48 & 438.63\\
17 & $7.87\times10^5$ & $5.20\times10^7$ & 132.29 & 13.00 & 3.16 & $6.02\times10^{-1}$ & $3.73\times10^{-1}$ & $1.68\times10^{-4}$ & 869.74 & 529.15\\
18 & $8.49\times10^5$ & $6.63\times10^7$ & 156.21 & 11.30 & 3.09 & $5.98\times10^{-1}$ & $3.58\times10^{-1}$ & $1.84\times10^{-4}$ & 888.09 & 531.35\\
19 & $7.31\times10^5$ & $4.94\times10^7$ & 134.99 & 13.17 & 3.14 & $6.02\times10^{-1}$ & $3.59\times10^{-1}$ & $1.85\times10^{-4}$ & 775.96 & 462.90\\
\hline
\end{tabular}
\caption{\label{tab1} Summary of small-world properties for the studied
  KKI graphs. $N$, $N_\text{edges}$: number of nodes and edges. $\langle k\rangle$: mean degree. $L$: average
  shortest path length. $L_r$: expectation
  value for the average shortest path length in Erd\H{o}s-R\'enyi
  graphs with the same $N$ and $N_\text{edges}$.
  $C^W$, $C^\Delta$: clustering coefficients defined by Eq.~\ref{Cws}
  and \ref{Cg}, respectively.
  $C_r$: mean clustering coefficient in Erd\H{o}s-R\'enyi graphs.
  $\sigma^W$, $\sigma^\Delta$: small-world coefficient defined by
  Eq.~\ref{swcoef}, based on either $C^W$ or $C^\Delta$.
}
\end{center}

\end{table*}

\section*{Conclusions}

Let us return to our introductory question: are the structural
connectome graphs from the OCP database scale-free, small-world networks?
As far as the adjective ``scale-free'' is concerned, the answer is
clearly no.
We have applied model selection based on the Akaike information criterion
to 10 graphs comparing 6 different degree distribution models.
The observed distributions are best fitted by the generalized Weibull
function with a stretched exponential tail $\propto\exp(-k^\beta)$.
Most of the exponents $\beta$ are between 0.2 and 0.5, which may hint
at a universal trend.
In some cases a truncated power law is a plausible alternative.
However, the truncation occurs at a degree much smaller than the number of
nodes in the network so that one cannot regard these distributions as
scale-free.

Unlike the term ``scale-free'', the adjective ``small-world'' does
apply to the OCP connectomes in the sense that the small-world
coefficients are much larger than 1.
We have performed a finite size scaling analysis using several graphs
and found no dependence between the number of nodes $N$ and the
small-world coefficients.
The average path length, however, decreases as $N$ increases.
The resolution to this apparent paradox is that the average degree
$\langle k\rangle$ increases with $N$ so that there is an
increasing number of shortcuts through the network.
On the other hand, we obtained small topological dimensions
characteristic of large-world networks.
The dimensions show a tendency to grow as the sizes of the studied
graphs increase.
The limit $N\to\infty$ here is taken by increasing $N$ for a
fixed voxel size.
The absence of a scale-free degree distribution suggests that this
limit may not be equivalent to fixing the brain volume and instead
resolving the details of the connectome at an infinitely small scale.
For this reason, it is difficult to judge whether Griffiths effects
can be found in the brain, but the small topological dimensions that
we have observed warrant further investigation.

Our analysis has been based on unweighted graphs.
More realistically, however, the connectome is a weighted, modular network.
Links between modules are known to be much weaker than the
intra-module connections.
Thus, future studies should take into account that signals in the
brain propagate on a weighted, heterogeneous network,
where generic slow dynamics is a distinct possibility~\cite{CCdyn}.
The methods presented here to characterize degree distributions and
topological dimensions can be generalized to the weighted case.
We hope that, as more precise and finely resolved connectome data will
become available, future research will be able to assess whether Griffiths
phases can indeed occur in the brain.

\section*{Acknowledgments}

We thank the staff of the Open Connectome project for help and discussions.
Support from the Hungarian research fund OTKA (Grant No. K109577) and
the European Commission (project number FP7-PEOPLE-2012-IEF 6-4564/2013)
is gratefully acknowledged.

\section*{Supplementary Information: \\
Justification of $\AIC_c$ as model selection criterion}
Among the various criteria for model selection that have been proposed in the
literature, the Akaike and Bayesian information criteria (commonly abbreviated
as AIC and BIC, respectively) are those that are most frequently
employed in the analysis of empirical data.
Both criteria have the general form
\begin{equation}
  -2\ln(\mathcal{L}(\hat{\mathbf{v}})) + aK, \tag{S1}
\end{equation}
where $\mathcal{L}$ is the likelihood function, $\hat{\mathbf{v}}$ the model
parameters that maximize $\mathcal{L}$, and $K$ is the number of parameters.
The difference between AIC and BIC is the prefactor $a$ in the last
term,
\begin{equation}
  a =
  \begin{cases}
    2 & \text{for AIC},\\
    \ln(N) & \text{for BIC}, 
  \end{cases}
\tag{S2} \label{penalty}
\end{equation}
where $N$ is the number of data points.
For $N\geq8$, the BIC gives thus a larger penalty than the AIC for
every additional parameter in the model.

BIC-based model selection is known to be consistent in the sense that
it almost surely selects the true model if it is among the candidates
and $N$ is infinitely large~\cite{Kim_etal12}.
AIC-based model selection is generally not consistent, but unlike the
BIC it is, in statistical parlance, efficient: in the limit
$N\to\infty$ the AIC selects a candidate model with the smallest
Kullback-Leibler divergence from the true model, even if the true
model is not among the candidates~\cite{Vrieze12, Aho_etal14}.
The true model is in most biological applications indeed unlikely to be included
in the candidate set because it usually contains a large number of unknown
parameters, which may furthermore increase with $N$.
The consistency of the BIC and the efficiency of the AIC are asymptotic
statements valid for $N\to\infty$.
While it is insightful to know under which conditions the AIC or BIC
are asymptotically optimal, these properties are ultimately of theoretical
rather than practical relevance for finite data sets.
The best guidance which criterion to choose comes from simulation studies
involving finite data.

With a random number generator we have created finite data that pose a similar
challenge to the model selection algorithm as the observed degree
distributions.
The data are $N=10^6$ random numbers, independently sampled from the distribution
\begin{align}
  &\Pr(k_\text{obs} > k) =
  \tag{S3} \label{Pr_power_test}\\
  &\hspace{.5cm}\begin{cases}
    \exp\left[-\beta(k_c+\alpha)^{-1}k\right] & \text{if $0 < k\leq
      k_c$,}\\
    C(k_c,\alpha,\beta) \times
    (k+\alpha)^{-\beta} & \text{if $k>k_c$,}
  \end{cases}\nonumber
\end{align}
where $k_\text{obs}$ is the generated random integer and the prefactor on the
second line is $C(k_c,\alpha,\beta) =
(k_c+\alpha)^\beta\exp\left[-\beta(k_c+\alpha)^{-1}k_c\right]$.
The tail of this distribution is a power law.
If $k$ were real-valued rather than an integer, the right-hand side of
Eq.~(\ref{Pr_power_test}) would be differentiable everywhere.
We have chosen this distribution for two reasons.
First, we want to test whether it is possible to determine with
information criteria that the true model is a
power-law (POW in Table~1).
Second, we want to find out which criterion is better at determining
the precise crossover point $k_c$ in the presence of random fluctuations.
The motivation is that, if the brain were indeed scale-free, we would
correctly identify the power-law model and its parameterization.
We fixed $\alpha=10$, $\beta=2.5$ and allowed $k_c$ to vary.

\renewcommand{\thefigure}{S\arabic{figure}}
\begin{figure}
  \includegraphics[height=5.5cm]{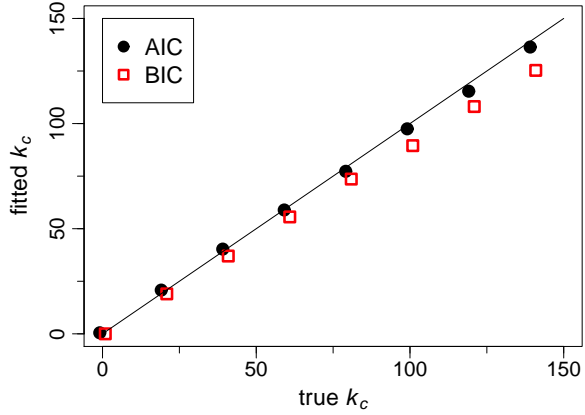}
  \caption{ \label{kc_mc}The true crossover point $k_c$ in
    Eq.~(\ref{Pr_power_test}) versus values fitted by selecting among
    the candidate models of Eq.~(\ref{Pr_power_cand}).
    The selected models either minimize AIC (black filled circles) or
    BIC (red open squares).
    The diagonal line indicates perfect prediction.
    Standard errors are smaller than symbol sizes.
    }
\end{figure}

For our first exploratory Monte Carlo simulations, we compared whether
AIC- or BIC-based model selection is more successful at determining
the true $k_c$ from the POW candidate models
\begin{align}
  &\text{Pr}_\text{candidate}(k_\text{obs} > k) =
  \tag{S4} \label{Pr_power_cand}\\
  &\hspace{0.5cm}\begin{cases}
    1-\sum_{i=1}^k A_i & \text{if $0<k\leq k_c$},\\
    (1-\sum_{i=1}^{k_c}A_i)\left(\frac{k_c+\alpha}{k+\alpha}\right)^\beta
    & \text{if $k>k_c$},
  \end{cases}\nonumber.
\end{align}
where $\alpha$, $\beta$ and $A_1, \ldots, A_{k_c}$ are adjustable parameters.
The crossover point $k_c$ is not itself a parameter in
Eq.~(\ref{Pr_power_cand}), but rather an index for different candidate
models.

We summarize the results in Fig.~S6.
The AIC reconstructs the crossover point almost accurately in the
entire range $0 \leq k_c \leq 150$.
The BIC underestimates $k_c$, especially when $k_c$ is large.
The reason lies in the BIC's steeper penalty for additional parameters
in Eq.~(\ref{penalty}).
Another obstacle for the BIC is that the true model (i.e.\ the
distribution of Eq.~\ref{Pr_power_test} with the three parameters
$\alpha$, $\beta$ and $k_c$) is generally not one of the candidates of
Eq.~(\ref{Pr_power_cand}) because those contain more parameters than necessary, in
particular one parameter $A_i$ for every $0<i\leq k_c$.
The AIC, by contrast, correctly chooses a candidate that still describes the input
best among the wrong (i.e.\ overparameterized) models: for $k>k_c$ the
distribution decays $\propto (k+\alpha)^{-\beta}$ whereas there is no
power law in the region $0<k\leq k_c$.

Having found that the AIC is better suited for model selection in the
present context, we investigate next whether the AIC reliably identifies
the correct model among those listed in Table~1.
We generate 10 independent sets of $N=10^6$ random numbers each.
These are drawn from the distribution of
Eq.~(\ref{Pr_power_test}) with $\alpha=10$, $\beta=2.5$, and now we also fix
$k_c=100$.
For 8 out of these 10 sets, the power law had the smallest AIC. In
the remaining two, the power law's AIC differed from the minimum only
by $\Delta_\text{POW}=0.05$ in
one case and $\Delta_\text{POW} = 0.30$ in the other case.
As we have explained after Eq.~(5),
such small $\Delta_\text{POW}$ are still interpreted
as substantial empirical support for the power law.
We conclude that, if the connectome's degree distribution were
scale-free, the AIC would have correctly identified a power law as a
plausible candidate model.

  observed network is not the full scale-free network, but only a
  random sample of 20\% of the edges.
  Such subsampling of the true network might mimic that the data are
  collected at a finite resolution and therefore only a fraction
  of the fiber tracts might be detected.
  For a fixed power-law degree distribution ($\alpha=10$, $\beta=2.5$,
  $N=2\cdot 10^6$, $k_c=0$), we first construct a
  corresponding network with the configuration model~\cite{MolloyReed95}.
  We then randomly select 20\% of the edges.
  The degrees of this subgraph have the distribution~\cite{Stumpf_etal05}
  \begin{equation}
    P(k_\text{obs} = j) = \sum_{i=j}^\infty \binom ij 0.2^j\,
    0.8^{i-j} P(k_\text{true}=i).
    \tag{S5}
  \end{equation}
  Here $k_\text{true}$ is the true degree of a node (i.e.\ in our
  model a power-law distribution POW as in Table 1 of the main text)
  and $k_\text{obs}$ is the degree observed in the sampled network.
  We removed all nodes with degree zero from the network to be
  consistent with the procedure used by the OCP.

  Applying AIC-based model selection with the candidates of Table 1,
  we find that the differences between the heavy-tailed
  distributions POW, LGN, TPW and GWB are smaller for the sampled
  network than for the full network.
  In ten test runs, we could always rule out EXP, WBL and in one case
  LGN because $\Delta>10$, but otherwise $\Delta$ was below this
  threshold.
  Still, POW had the smallest $\AIC$ in eight out of ten cases and
  came in a close second in the remaining two.
  In summary, AIC-based model selection would have correctly included
  the power law in the set of plausible candidates even if the
  observed network had been a small sample of a network with a
  scale-free degree distribution.

There is one fine detail to note when the dimension $K$ of the
model approaches the sample size $N$.
In such cases the AIC is a negatively biased estimate of the
Kullback-Leibler information~\cite{HurvichTsai89}.
An approximate correction for this bias is the
additional penalty term $\frac{2K(K+1)}{N-K-1}$ in Eq.~(4).
This modified version of the AIC is commonly abbreviated as $\AIC_c$.
Although the difference is small in the present context, it has been
suggested that $\AIC_c$ should routinely be used instead of the
AIC~\cite{HurvichTsai89, BurnhamAnderson01}.
We have followed this advice throughout this article.

\end{document}